\documentclass[conference]{IEEEtran}
\IEEEoverridecommandlockouts
\usepackage{cite}
\usepackage{amsmath,amssymb,amsfonts}
\usepackage{algorithmic}
\usepackage{graphicx}
\usepackage{geometry}
\usepackage{textcomp}
\usepackage{xcolor}
\usepackage{booktabs,multirow}
\usepackage{makecell}
\usepackage{subfigure}
\usepackage{float}
\def\BibTeX{{\rm B\kern-.05em{\sc i\kern-.025em b}\kern-.08em
    T\kern-.1667em\lower.7ex\hbox{E}\kern-.125emX}}
\begin{document}

\title{Integrated Sensing and Communication enabled Multiple Base Stations Cooperative UAV Detection \\
}

\author{\IEEEauthorblockN{Xi Lu$^1$, Zhiqing Wei$^{1,*}$, Ruizhong Xu$^1$, Lin Wang$^1$, Bohao Lu$^1$, Jinghui Piao$^1$}
	\textit{Key Laboratory of Universal Wireless Communications, Ministry of Education,}\\
	$^1$\text{Beijing University of Posts and Telecommunications, Beijing 100876, China}\\
	$^1$Email: \{luxi, weizhiqing, xuruizhong, wlwl, bohaolu, piaojinghui\}@bupt.edu.cn \\}

\maketitle

\begin{abstract}
Integrated sensing and communication (ISAC) exhibits notable potential for sensing the unmanned aerial vehicles (UAVs), facilitating real-time monitoring of UAVs for security insurance. Due to the low sensing accuracy of single base stations (BSs), a cooperative UAV sensing method by multi-BS is proposed in this paper to achieve high-accuracy sensing. Specifically, a multiple signal classification (MUSIC)-based symbol-level fusion method is proposed for UAV localization and velocity estimation, consisting of a single-BS preprocessing step and a lattice points searching step. The preprocessing procedure enhances the single-BS accuracy by superposing multiple spectral functions, thereby establishing a reference value for subsequent lattice points searching. Furthermore, the lattice point with minimal error compared to the preprocessing results is determined as the fusion result. Extensive simulation results reveal that the proposed symbol-level fusion method outperforms the benchmarking methods in localization and velocity estimation.
\end{abstract}


\section{Introduction}

As a promising technology for the sixth generation (6G) networks, integrated sensing and communication (ISAC) enables the incorporation of wireless sensing capabilities into wireless communication networks. Unmanned aerial vehicles (UAVs) have attracted widespread attention due to their flexibility and low cost, whereas giving rise to security concerns, such as the unauthorized UAVs posing potential hazards. To address this problem, monitoring the real-time status of UAVs is indispensable. Thus, ISAC-enabled mobile communication networks can be utilized to achieve high-precision and wide-area sensing of UAVs, which is more cost-effective than traditional radar systems~\cite{cao2022cellular,lou2022uav,zhao2019radar}.

Some research elaborates on ISAC-enabled UAV sensing. Zhao \emph{et al}.~\cite{zhao2019radar} proposed a UAV detection algorithm based on 5G millimeter wave radar, acquiring status information of UAVs. Yang \emph{et al}.~\cite{yang2016passive} employed a passive radar network featuring orthogonal frequency division multiplexing (OFDM) signals to detect and differentiate UAVs from other targets. Ai \emph{et al}.~\cite{ai2021passive} utilized 5G signals for UAV detection, obtaining range-Doppler tracking information of UAVs. Most of the existing research focuses on single base station (BS) sensing. However, single-BS sensing is constrained by accuracy, particularly for targets at the edge of the sensing range. Simultaneously, the small radar cross section (RCS) of UAVs reduces the signal-to-noise ratio (SNR) of received echo signals, which affects the sensing performance of single-BS\cite{cao2022cellular}. Thus, it is significant to investigate the multi-BS cooperative sensing to enhance sensing performance~\cite{wei2023symbol}. Research in multi-point cooperative sensing originates from radar theory, encompassing data-level and signal-level fusion techniques.

Data-level fusion method integrates estimated results from individual stations, with the process typically categorized into weighted and maximum likelihood methods. As for the weighted method, Ren \emph{et al}.~\cite{ren2021improved} devised a strategy to determine the weight of each radar by considering factors such as consistency, stability, and correlation between parameters. Nuss \emph{et al}.~\cite{nuss20163d} utilized multiple input multiple output (MIMO)-OFDM radars for multiple targets sensing, superposing measurement matrices from each radar. Zhang \emph{et al}.~\cite{zhang2021efficient} proposed a fusion algorithm for multi-target active and passive sensing, where the estimated results are combined based on their relative amplitudes. As for the maximum likelihood method, Gao \emph{et al}.~\cite{gao2021reliable} introduced a grid-based probabilistic fusion framework to mitigate the influence of radar ghosts. Some research proposed fusion methods based on multi-radar active or passive sensing. These methods define a maximum likelihood function for each position grid, with the estimation result being the position corresponding to the maximum value of the function\cite{weiss2011direct,yang2017novel,weiss2009direct,weiss2004direct}. Although data-level fusion is computationally efficient, it inevitably involves information loss since each station processes signals before fusion.

Signal-level fusion method directly superposes signals from multiple stations to enhance signal-to-noise ratio (SNR). Yang \emph{et al}.~\cite{yang2013phase} introduced a phase difference estimation method for distributed coherent aperture radar, effectively improving estimation accuracy and overall radar network performance. Compared with data-level fusion, signal-level fusion reduces information loss but needs stringent time and phase synchronization. Considering the characteristics of data-level and signal-level fusion, Wei \emph{et al}.~\cite{wei2023symbol} proposed a symbol-level fusion method, which extracts amplitude and phase information of single-BS for fusion. In comparison to the fusion algorithm above, symbol-level fusion minimizes information loss and relaxes the demands on synchronization accuracy.

In this paper, we propose a symbol-level active sensing fusion method in UAV localization and velocity estimation. Specifically, the multiple signal classification (MUSIC) algorithm is utilized to extract the spectral function and noise subspace of angle, distance, and velocity estimation for fusion, instead of the estimation results. Then, the spectral functions are superposed to improve the precision of single-BS sensing. Moreover, a lattice points searching method is proposed to achieve the symbol-level fusion of multi-BS. The main contributions of this paper are listed as follows.
\begin{itemize}
 	
 	\item The MUSIC algorithm is utilized for angle, distance, and velocity estimation. Furthermore, the preprocessing algorithm is proposed, which superposes multiple spectral functions to diminish noise and improve accuracy for single-BS, providing more accurate estimation ranges and utilized as reference values for subsequent fusion methods.
 	\item The symbol-level fusion method based on the combination of lattice points searching and error accumulation is proposed\cite{wei2023symbol}, where the point with the minimum error is identified as the result, quantifying the error through the orthogonality between the noise subspace of preprocessing and the parameter vectors of lattice point.
 	\item Extensive simulations demonstrate the superiority of the proposed symbol-level fusion method. In comparison to the benchmarking methods, the proposed approach exhibits reduced root mean square error (RMSE) in UAV localization and velocity estimation.
 	
\end{itemize}

\section{System and Signal Model}

In this section, the multiple ISAC BSs cooperative sensing model and the signal model are introduced. The received signal model serves as the prerequisite for subsequent localization, velocity estimation, and fusion methods.

\subsection{System Model}

\begin{figure}[!h]
	\centering
	\includegraphics[width=0.65\linewidth]{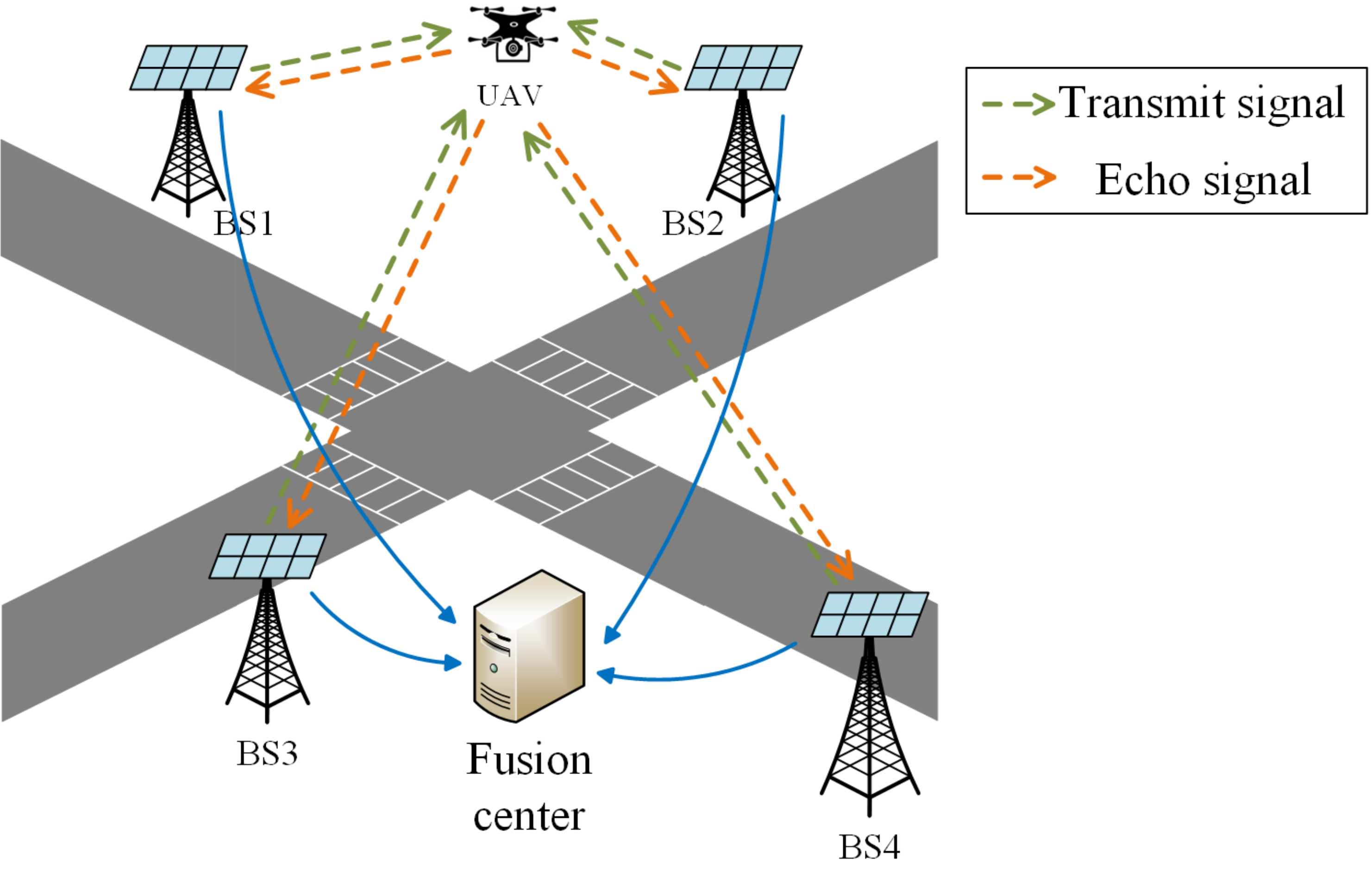}
	\caption{Multiple ISAC BSs cooperative sensing.}
	\label{fig:Scenario}
\end{figure}

This paper consider a multiple ISAC BSs cooperative sensing scenario, as shown in Fig.~\ref{fig:Scenario}. Each BS is equipped with well-separated transmit uniform planar array (UPA) and receive UPA. The BS transmits downlink signals while receiving echo signals reflected from UAV for sensing. The position of UAV is denoted as $({x_{t}},{y_{t}},{z_{t}})$, where ${x_{t}},{y_{t}},$ and ${z_{t}}$ are the three-dimensional coordinates. The velocity is denoted as $({v_{t}},{\theta _{t}},{\varphi _{t}})$, where ${v_{t}}$ is the magnitude of velocity while ${\theta _{t}}$ and ${\varphi _{t}}$ represent the azimuth and elevation angles of the direction of velocity, respectively.

\subsection{Transmit Signal Model} 

The OFDM signal transmitted by the $n$-th BS is given by
\begin{equation}
	{s_{T,n}}(t) = \sum\limits_{\mu  = 0}^{{N_s} - 1}  \sum\limits_{m = 0}^{{N_c} - 1}  {d_{m,\mu }}{e^{j(2\pi ({f_0} + m\Delta f)t + {\varphi _0})}}{\rm{rect}}(\frac{{t - \mu {T_s}}}{{{T_s}}}),
	\label{eq:ofdm_transmit}
\end{equation}
where ${N_s}$ and ${N_c}$ represent the number of OFDM symbols and the number of subcarriers, respectively; ${d_{m,\mu}}$ is the symbol transmitted at the $m$-th subcarrier of the $\mu$-th OFDM symbol, ${f_0}$ represents the carrier frequency, $\Delta f$ is the subcarrier spacing, ${\varphi_0}$ is the initial phase; ${T_s}$ stands for the total duration of OFDM symbol, which is the sum of the duration of cyclic prefix (CP) and the duration of OFDM symbol without CP; and $\rm{rect}(*)$ is the rectangular function, which takes the value of 1 during the duration of each symbol and 0 for others~\cite{sturm2011waveform}.

\subsection{Received Echo Signal Model}

\begin{figure}[!h]
	\centering
	\includegraphics[width=0.45\linewidth]{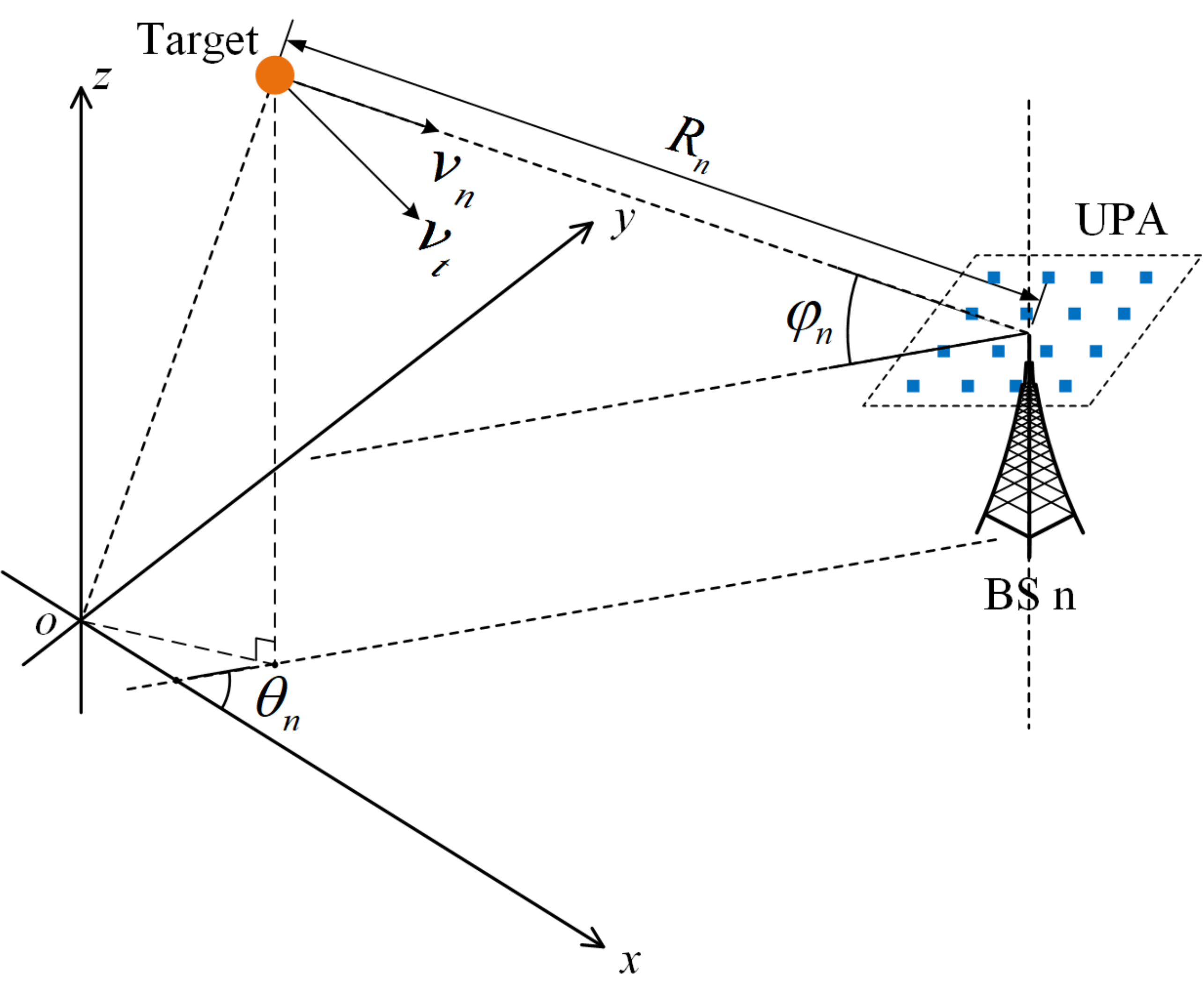}
	\caption{Parameters of single-BS sensing.}
	\label{fig:single-BSn}
\end{figure}

The echo signal reflected by the target undergoes a phase shift attributed to time delay and  Doppler shift caused by the motion of the UAV. Additionally, environmental factors bring Gaussian white noise and amplitude attenuation. To mitigate the influence of the transmitted communication signal, the echo signal is divided by the transmitted data to obtain the channel information~\cite{sturm2011waveform}. The expression for the $\mu$-th symbol at the $m$-th subcarrier of the $n$-th BS is
\begin{equation}
	{{\bf{C}}_n}(m,\mu ) = {U_n}{e^{ - j2\pi m\Delta f\frac{{2{R_{n}}}}{c}}}{e^{j2\pi {f_0}\frac{{2{v_{n}}}}{c}\mu {T_s}}} + \eta (m,\mu ),
	\label{eq:ofdm_echo}
\end{equation}
where ${R_{n}}$ and ${v_{n}}$ are the distance and radial velocity of UAV at the direction of the $n$-th BS, as depicted in Fig.~\ref{fig:single-BSn}. $U_n$ represents the amplitude attenuation, and $\eta$ denotes the noise, which follows a Gaussian distribution with a mean of 0 and a variance of ${\sigma ^2}$.

As shown in Fig.~\ref{fig:single-BSn}, each BS is equipped with a receive UPA with the size of $M_{r} \times M_{c}$, placed parallel to the \emph{xoy} plane. ${\theta _{n}}$ and ${\varphi _{n}}$ represent the azimuth and elevation angles of the received signal, respectively. The range of ${\theta _{n}}$ spans from $-\pi$ to $\pi$, with the positive direction anticlockwise from the positive $x$-axis. Similarly, the range of ${\varphi _{n}}$ spans from $- \pi/2$ to $\pi/2$, with the positive direction anticlockwise from the plane parallel to the \emph{xoy} plane.

It is assumed that the UAV is located above BSs, with the elevation angle ${\varphi _{n}}$ constrained within the range of $-\frac{\pi }{2} < {\varphi_{n}} < 0$. Considering the top-left element of the antenna array as the reference element, the azimuth and elevation angles influence the phase difference between antenna elements. The phase difference between the antenna element at the $f$-th row and the $g$-th column and the reference element can be expressed as the steering vector, which is

\begin{equation}
	\begin{array}{c}
		\begin{aligned}
			{{\bf{k}}_{a,n}}(f,g) &= {e^{jk2\pi \frac{{\Delta {d_{n(f,g)}}}}{\lambda }}}\\
			&= {e^{jk2\pi \frac{{(f - 1)\Delta x\cos \theta \cos \varphi  + (g - 1)\Delta y\sin \theta \cos \varphi }}{\lambda }}},
		\end{aligned} \\
			
		\begin{small}
			\left\{ {\begin{array}{*{20}{l}}
					{k = 1,\theta  = {\theta _n},\varphi  = {\varphi _n}},&{0 \le {\theta _n} < \frac{\pi }{2}}\\
					{k =  - 1,\theta  = \pi  - {\theta _n},\varphi  = \pi  - {\varphi _n}},&{\frac{\pi }{2} \le {\theta _n} < \pi }\\
					{k =  - 1,\theta  = \pi  + {\theta _n},\varphi  = \pi  + {\varphi _n}},&{ - \pi  \le {\theta _n} <  - \frac{\pi }{2}}\\
					{k = 1,\theta  =  - {\theta _n},\varphi  =  - {\varphi _n}},&{ - \frac{\pi }{2} \le {\theta _n} < 0}
			\end{array}} \right.
		\end{small}
	\end{array}
	\label{eq:ka}
\end{equation}
where $\lambda$ is wavelength, $\Delta d_{n(f,g)}$ is the distance difference between antenna elements along the direction of echo signal. $\Delta x$ and $\Delta y$ are the spacing between columns and rows of the antenna array, respectively, both of which are set as $\lambda/2$ in this paper. As shown in (\ref{eq:ofdm_echo}), the echo signal ${\bf{C}}_n$ is a matrix with dimensions $N_{c} \times N_{s}$. The received signal at different BSs is represented as

\begin{equation}
	{{\bf{Y}}_{a,r,n}} = {{\bf{k}}_{a,n}}{{\bf{s}}_{r,n}} + {{\bf{N}}_{a,n}},
	\label{eq:received signal}
\end{equation}
where ${\bf{s}}_{r,n}$ is the $r$-th row of echo signal matrix ${\bf{C}}_n$, ${\bf{N}}_{a,n}$ represents the noise. ${\bf{k}}_{a,n}$ is the receive steering vector, which can be expressed as

\begin{equation}
	{{\bf{k}}_{a,n}}\!\! =\!\!{\left( {\begin{array}{*{20}{c}}
				{{\! e^{jk2\pi \frac{{\Delta {d_{n(1,1)}}}}{\lambda }}}}\!\!&\!\!{{e^{jk2\pi \frac{{\Delta {d_{n(2,1)}}}}{\lambda }}}}\!\!&\!\!{ \cdots }\!\!&\!\!{{e^{jk2\pi \frac{{\Delta {d_{n({M_r},{M_c})}}}}{\lambda }}}\!}
		\end{array}} \right)^T}.
	\label{direction vector}
\end{equation}

\section{Single-BS Estimation Preprocessing}
In this section, MUSIC algorithm is utilized to estimate the azimuth and elevation angles, distance, and radial velocity between target UAV and BS. The preprocessing procedure involving the superposition of multiple spectral functions is proposed to enhance the accuracy of single-BS.

\subsection{Azimuth and Elevation Angles Estimation}

The received signal matrix ${\bf{Y}}_{a,r,n}$ is of size ${M_r}{M_c} \times {N_s}$, the auto-correlation matrix of which is calculated as

\begin{equation}
	{{\bf{R}}_{a}} = \frac{1}{{{N_c}{N_s}}}{{\bf{Y}}_{a,r,n}}{{\bf{Y}}_{a,r,n}}^H.
	\label{eq:auto-correlation}
\end{equation}

Applying eigenvalue decomposition to ${{\bf{R}}_{a}}$, the noise subspace ${{\bf{U}}_{a,r,n}} = {{\bf{U}}_x}(\ \!:\ \!,\ 2\!:\!{M_r} \times {M_c})$ is constructed, in which ${{\bf{U}}_x}$ represents the orthogonal eigen matrix. The spectral function is obtained as\cite{chen2023multiple}

\begin{equation}
	{P_{a,r,n}({\theta _{l}},{\varphi _{l}})} = {{\bf{k}}_{a,n}^H({\theta _l},{\varphi _l}){{\bf{U}}_{a,r,n}}{\bf{U}}_{a,r,n}^H{{\bf{k}}_{a,n}}({\theta _l},{\varphi _l})}.
	\label{eq:spectral function}
\end{equation}

The estimated elevation and azimuth angles can be obtained by searching the minimum value in (\ref{eq:spectral function}). However, relying solely on one row of ${\bf{C}}_n$ may not yield sufficiently accurate results due to the influence of noise. We superpose the spectral function of ${N_r}$ rows of ${\bf{C}}_n$ to mitigate the influence of random noise. Then, the angle estimation is equivalent to searching the peak of the reciprocal of superposed spectral function, which is

\begin{equation}
	\begin{aligned}
		{P_{a,n}({\theta _{l}},{\varphi _{l}})} 
		&\!=\! \frac{1}{{\sum\limits_{r = 1}^{{N_r}} {{\bf{k}}_{a,n}^H({\theta _l},{\varphi _l}){{\bf{U}}_{a,r,n}}{\bf{U}}_{a,r,n}^H{{\bf{k}}_{a,n}}({\theta _l},{\varphi _l})} }} \\
		&\!=\! \frac{1}{{{\bf{k}}_{a,n}^H({\theta _l},{\varphi _l})\sum\limits_{r = 1}^{{N_r}} {{{\bf{U}}_{a,r,n}}{\bf{U}}_{a,r,n}^H} {{\bf{k}}_{a,n}}({\theta _l},{\varphi _l})}}.
		\label{eq:Psum function}
	\end{aligned}
\end{equation}

By searching the peak of (\ref{eq:Psum function}), the high-precision estimation results for the azimuth and elevation angles are denoted as ${\tilde \theta _{n}}$ and ${\tilde \varphi _{n}}$, respectively.

\subsection{Distance and Velocity Estimation}
As for distance and velocity, we first extract one column or row of ${\bf{C}}_n$ for estimation, respectively. The noise subspace ${{{\bf{U}}_{d,c,n}}}$ and ${{{\bf{U}}_{v,c,n}}}$ for distance and velocity estimation can be calculated by the $c$-th column and row, which are respectively of size ${N_c} \times ({N_c}-1)$ and ${N_s} \times ({N_s}-1)$. Subsequently, distance and velocity vectors are similar to the structure of steering vector in (\ref{direction vector}), with the search value ${R_l}$ and ${v_l}$, which can be expressed as

\begin{equation}
	{{\bf{k}}_{d,n}} \!=\! \left(\! {\begin{array}{*{20}{c}}
			1\!&\!{{e^{ - j2\pi \Delta f\frac{{2{R_l}}}{c}}}}& \!\cdots\! &\!{{e^{ - j2\pi ({N_c} - 1)\Delta f\frac{{2{R_l}}}{c}}}}\!
	\end{array}}\! \right)^T,
	\label{eq:distance vector}
\end{equation}

\begin{equation}
	{{\bf{k}}_{v,n}} \!=\! \left( {\begin{array}{*{20}{c}}
			1\!&\!{{e^{j2\pi {f_0}\frac{{2{v_l}}}{c}{T_s}}}}& \!\cdots\! &{{e^{j2\pi ({N_s} - 1){f_0}\frac{{2{v_l}}}{c}{T_s}}}}\!
	\end{array}} \right)^T.
	\label{eq:velocity vector}
\end{equation}

Similar to the procedure for angle estimation, we superpose the spectral function of ${N_d}$ columns and rows of ${\bf{C}}_n$ for high-precision estimation, which is given by

\begin{equation}
	{P_{d,n}}({R_l}) = \frac{1}{{{\bf{k}}_{d,n}^H({R_l})\sum\limits_{c = 1}^{{N_d}} {{{\bf{U}}_{d,c,n}}{\bf{U}}_{d,c,n}^H} {{\bf{k}}_{d,n}}({R_l})}},
	\label{eq:distance spectral function}
\end{equation}

\begin{equation}
	{P_{v,n}}({v_l}) = \frac{1}{{{\bf{k}}_{v,n}^H({v_l})\sum\limits_{c = 1}^{{N_d}} {{{\bf{U}}_{v,c,n}}{\bf{U}}_{v,c,n}^H} {{\bf{k}}_{v,n}}({v_l})}}.
	\label{eq:velocity spectral function}
\end{equation}

By searching the peaks of (\ref{eq:distance spectral function}) and (\ref{eq:velocity spectral function}), the high-precision estimation results ${{\tilde R}_n}$ and ${\tilde v_{n}}$ can be obtained\cite{chen2023multiple}.

\subsection{UAV Localization and Velocity Estimation}\label{sec:results}
As for the localization, given ${{\tilde R}_n}$, ${\tilde \theta _{n}}$, ${\tilde \varphi _{n}}$ and the known positions of the BSs, each BS can calculate a set of localization results for the UAV, which is denoted as $({\tilde x_{t,n}},{\tilde y_{t,n}},{\tilde z_{t,n}})$.

\begin{figure}[!h]
	\centering
	\includegraphics[width=0.48\linewidth]{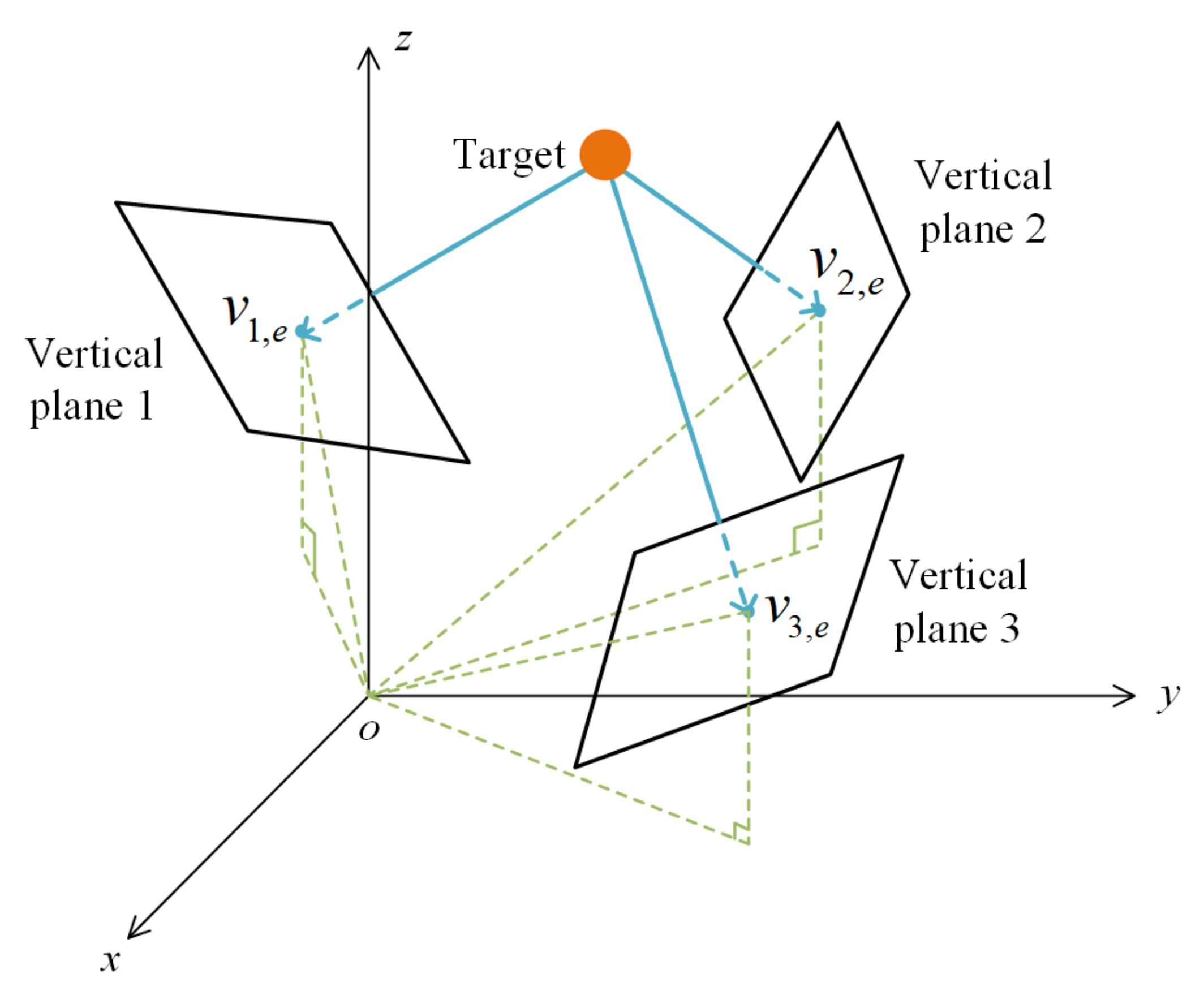}
	\caption{Velocity estimation of UAV.}
	\label{fig:UAV Velocity}
\end{figure}

As for the velocity estimation, three BSs can collectively determine one set of magnitude and direction estimations. The terminal of the velocity vector is situated on a plane perpendicular to the radial velocity vector of each BS. The intersection point of three vertical planes represents the result, as depicted in Fig.~\ref{fig:UAV Velocity}. Consequently, four sets of velocity estimation results $({\tilde v_{t,n}},{\tilde \theta _{t,n}},{\tilde \varphi _{t,n}})$ can be obtained, including magnitude and direction.

\section{Multi-BS Cooperative Sensing}

In this section, a lattice points searching fusion method based on error accumulation is employed for more precise estimations. The central point of the searching grid is determined by the average value derived from the estimation results of each BS.

\subsection{Multi-BS Cooperative Localization of UAV }\label{sec:localization fusion}

\begin{figure}[!h]
	\centering
	\includegraphics[width=0.45\linewidth]{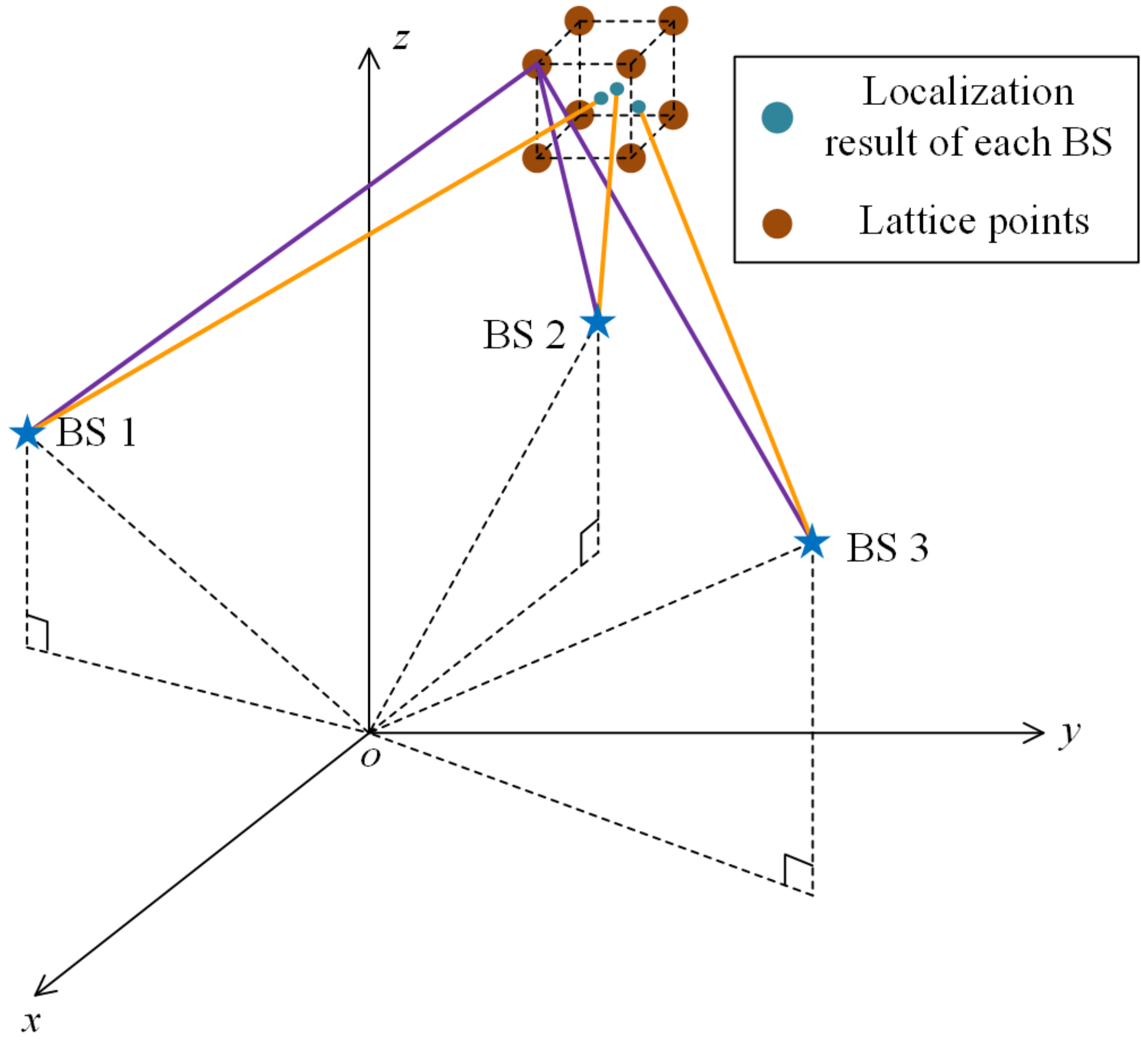}
	\caption{The lattice points for cooperative localization of UAV.}
	\label{fig:UAV position fusion}
\end{figure}

As illustrated in Fig.~\ref{fig:UAV position fusion}, the lattice points for the position of UAV are arranged as a cube. The radial distance ${R_{n,q}}$, azimuth angle ${\theta_{n,q}}$, and elevation angle ${\varphi_ {n,q}}$ in the direction of the $n$-th BS at lattice point $q$ are calculated, with the distance difference between antenna elements expressed as ${\Delta {d_{nq(f,g)}}}$.

The distance and direction vectors are represented as

\begin{equation}
	{{\bf{k}}_{d,n,q}} = \left(1\ \ {e^{ - j2\pi \Delta f\frac{{2{R_{n,q}}}}{c}}}\ \   \cdots \ \   {e^{ - j2\pi ({N_c} - 1)\Delta f\frac{{2{R_{n,q}}}}{c}}}\right)^T,
\end{equation}

\begin{equation}
	\begin{small}
		\begin{aligned}
			{{\bf{k}}_{a,n,q}}\!\! =\!\! \left(\!{e^{ jk2\pi \frac{{\Delta {d_{nq(1,1)}}}}{\lambda }}}\ {e^{ jk2\pi \frac{{\Delta {d_{nq(2,1)}}}}{\lambda }}}\ \! \cdots \! \ {e^{ jk2\pi \frac{{\Delta {d_{nq(M_{r},M_{c})}}}}{\lambda }}}\!\right)^T.
		\end{aligned}
	\end{small}
\end{equation}

The orthogonality between ${\bf{k}}_{d,n,q}$ and noise subspace ${{\bf{G}}_{d,n}} = \sum\limits_{c = 1}^{{N_d}} {{{\bf{U}}_{d,c,n}}{\bf{U}}_{d,c,n}^H} $ is calculated as the error between ${R_{n,q}}$ and high-precision distance estimation result ${\tilde R _{n}}$. Similarly, the orthogonality between ${\bf{k}}_{a,n,q}$ and noise subspace ${{\bf{G}}_{a,n}} = \sum\limits_{r = 1}^{{N_r}} {{{\bf{U}}_{a,r,n}}{\bf{U}}_{a,r,n}^H} $ is calculated as the error between ${\theta_{n,q}}$ and ${\tilde \theta _{n}}$, as well as ${\varphi_{n,q}}$ and ${\tilde \varphi_{n}}$. The orthogonality is calculated as

\begin{equation}
		{P_{d,n,q}}({R_{n,q}}) = \frac{1}{{{{\bf{k}}_{d,n,q}}^H({R_{n,q}}){{\bf{G}}_{d,n}}{{\bf{k}}_{d,n,q}}({R_{n,q}})}},
\end{equation}

\begin{equation}
		{P_{a,n,q}}({\theta _{n,q}},{\varphi _{n,q}})\! =\! \frac{1}{{{\bf{k}}_{a,n,q}^H({\theta _{n,q}},{\varphi _{n,q}}){{\bf{G}}_{a,n}}{{\bf{k}}_{a,n,q}}({\theta _{n,q}},{\varphi _{n,q}})}}.
\end{equation}

We sum $P_{r,n,q}({R_{n,q}})$ and $P_{a,n,q}({\theta _{n,q}},{\varphi _{n,q}})$ of $N_{f}$ BSs as the weight of point $q$, which is given by

\begin{equation}
	{W_{l,q}} = \sum\limits_{n = 1}^{{N_{f}}} {{P_{d,n,q}}({R_{n,q}})} + {P_{a,n,q}}({\theta _{n,q}},{\varphi _{n,q}}) .
\end{equation}

The lattice point with the maximum value of $W_{l,q}$ is the final estimation result $({x_{t,f}},{y_{t,f}},{z_{t,f}})$ of UAV localization.

\subsection{Multi-BS Cooperative Velocity Estimation of UAV}

\begin{figure}[!h]
	\centering
	\includegraphics[width=0.5\linewidth]{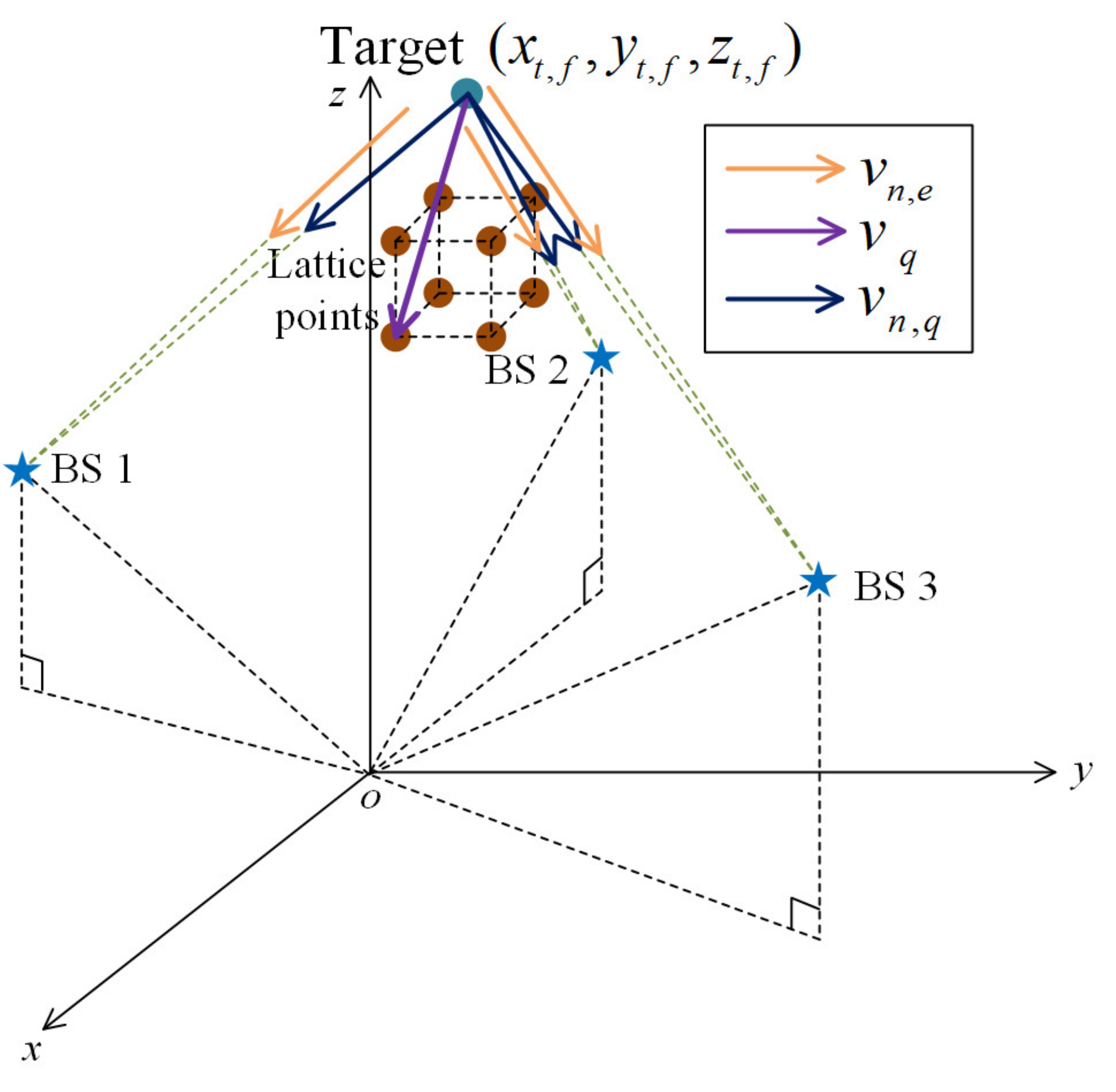}
	\caption{The lattice points for cooperative velocity estimation of UAV.}
	\label{fig:UAV velocity fusion}
\end{figure}

The lattice points for the terminal of the velocity vector are organized in the form of a cube, as depicted in Fig.~\ref{fig:UAV velocity fusion}. For a given lattice point $q$, the radial velocity $v_{n,q}$ in the direction of the $n$-th BS is computed.

The velocity vector can be expressed as

\begin{equation}
		{{\bf{k}}_{v,n,q}} = \left(1\ \ {e^{j2\pi {f_0}\frac{{2{v_{n,q}}}}{c}{T_s}}}\ \  \cdots \ \ {e^{j2\pi ({N_s} - 1){f_0}\frac{{2{v_{n,q}}}}{c}{T_s}}}\right)^T.
\end{equation}

Calculate the orthogonality between ${\bf{k}}_{v,n,q}$ and noise subspace ${{\bf{G}}_{v,n}} = \sum\limits_{c = 1}^{{N_d}} {{{\bf{U}}_{v,c,n}}{\bf{U}}_{v,c,n}^H} $ as the error between ${v_{n,q}}$ and ${\tilde v_{n}}$, which can be expressed as

\begin{equation}
	{P_{v,n,q}}({v_{n,q}}) = \frac{1}{{{{\bf{k}}_{v,n}}^H({v_{n,q}}){{\bf{G}}_{v,n}}{{\bf{k}}_{v,n}}({v_{n,q}})}}.
\end{equation}

We sum $P_{v,n,q}({v_{n,q}})$ of $N_{f}$ BSs as the weight of point $q$, which is given by

\begin{equation}
	{W_{v,q}} = \sum\limits_{n = 1}^{{N_{f}}} {{P_{v,n,q}}}.
\end{equation}

The lattice point with the maximum value of $W_{v,q}$ is the fusion result of the velocity vector of the UAV, from which the magnitude and direction of velocity $({v_{t,f}},{\theta _{t,f}},{\varphi _{t,f}})$ can be obtained.

\section{Simulation Results} 

The simulation parameters are listed in Table \ref{tab:parameters} unless otherwise stated. The fusion algorithm undergoes 300 Monte Carlo simulations. Given that 3 BSs contribute to the velocity estimation, a minimum number of 4 BSs is essential. The chosen 4 BSs are strategically positioned in disparate directions with the two-dimensional coordinates $(80,50)$, $(-30,85)$, $(40,-60)$, and $(-10,-70)$ and a height of 20 meters. The comparison encompasses RMSE of localization, involving the error in the $x$, $y$, $z$ axis, and RMSE of velocity estimation, including magnitude and direction errors.

\begin{table}[htbp]
	\caption{Simulation Parameters}
	\renewcommand\arraystretch{1.2}
	\begin{center}
		\begin{tabular}{ccc}
			\toprule
			\textbf{Parameter} & \textbf{Meaning} & \textbf{Value} \\
			\midrule
			$M_{r} \times M_{c}$ & Size of antenna array & $4 \times 4$ \\
			$N_{c}$ & Number of subcarriers & 128 \\
			$N_{s}$ & Number of symbols & 256 \\
			$f_{0}$ & Carrier frequency &24 GHz \\
			$\Delta f$ & Subcarrier spacing & 240 kHz \\
			$T_{s}$ & Duration of OFDM symbol & 5.208 $\mu$s \\
			$N_{BS}$ & Number of BSs & 4 \\
			$N_{f}$ & Number of fusion BSs & 3 $\sim$ 4 \\
			$({x_{t}},{y_{t}},{z_{t}})$ & Position of UAV & \makecell[c]{(-7.2873,6.8487,\\39.2624)m} \\
			$({v_{t}},{\theta _{t}},{\varphi _{t}})$ & Velocity of UAV & (23 m/s,$70^{\circ}$,$-40^{\circ}$) \\
			\bottomrule
		\end{tabular}
		\label{tab:parameters}
	\end{center}
\end{table}

Moreover, extensive simulations are conducted to compare the sensing performance of the proposed fusion method with the average value fusion method, data-level fusion method, and a single set of localization and velocity estimation results. The weight of lattice points in data-level fusion is determined by the absolute value between the results of preprocessing and the parameter values of each lattice point, which is given by

\begin{equation}
	\begin{small}
		{W_{l,q'}} = \sum\limits_{n = 1}^{{N_f}} {\left| {{{\tilde R}_n} - {R_{n,q}}} \right| + \left| {{{\tilde \theta }_n} - {\theta _{n,q}}} \right| + \left| {{{\tilde \varphi }_n} - {\varphi _{n,q}}} \right|}, 
	\end{small}
\end{equation}

\begin{equation}
	\begin{small}
		{W_{v,q'}} = \sum\limits_{n = 1}^{{N_f}} {\left| {{{\tilde v}_n} - {v_{n,q}}} \right|}.  
	\end{small}
\end{equation}

\subsection{Simulation Results of Single-BS Estimation}

\begin{figure}[!h]
	\centering  
	\subfigbottomskip=2pt 
	\subfigcapskip=-5pt 
	\subfigure[Radial distance estimation.]{
		\includegraphics[width=0.4\linewidth]{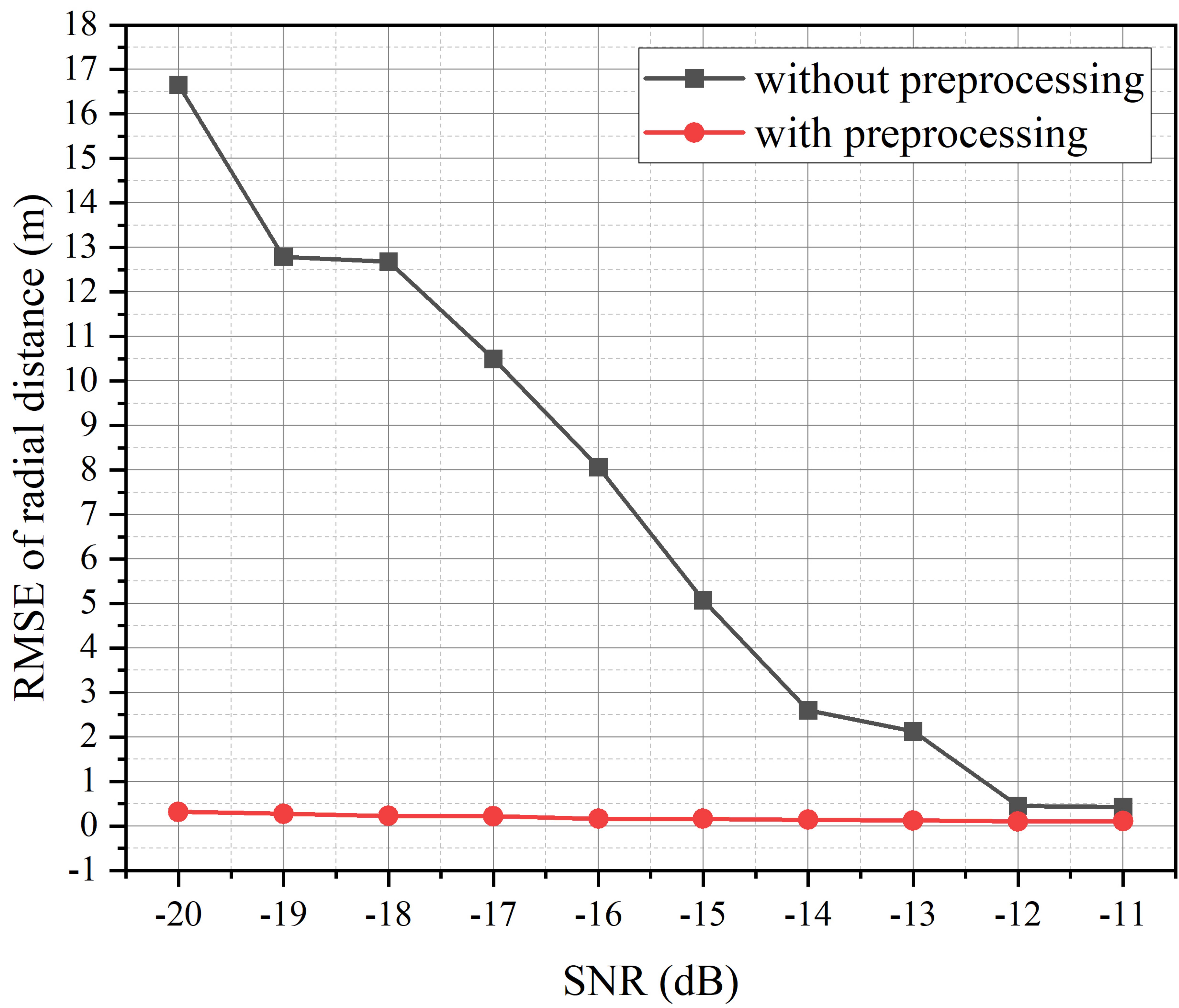}}
	\subfigure[Radial velocity estimation.]{
		\includegraphics[width=0.4\linewidth]{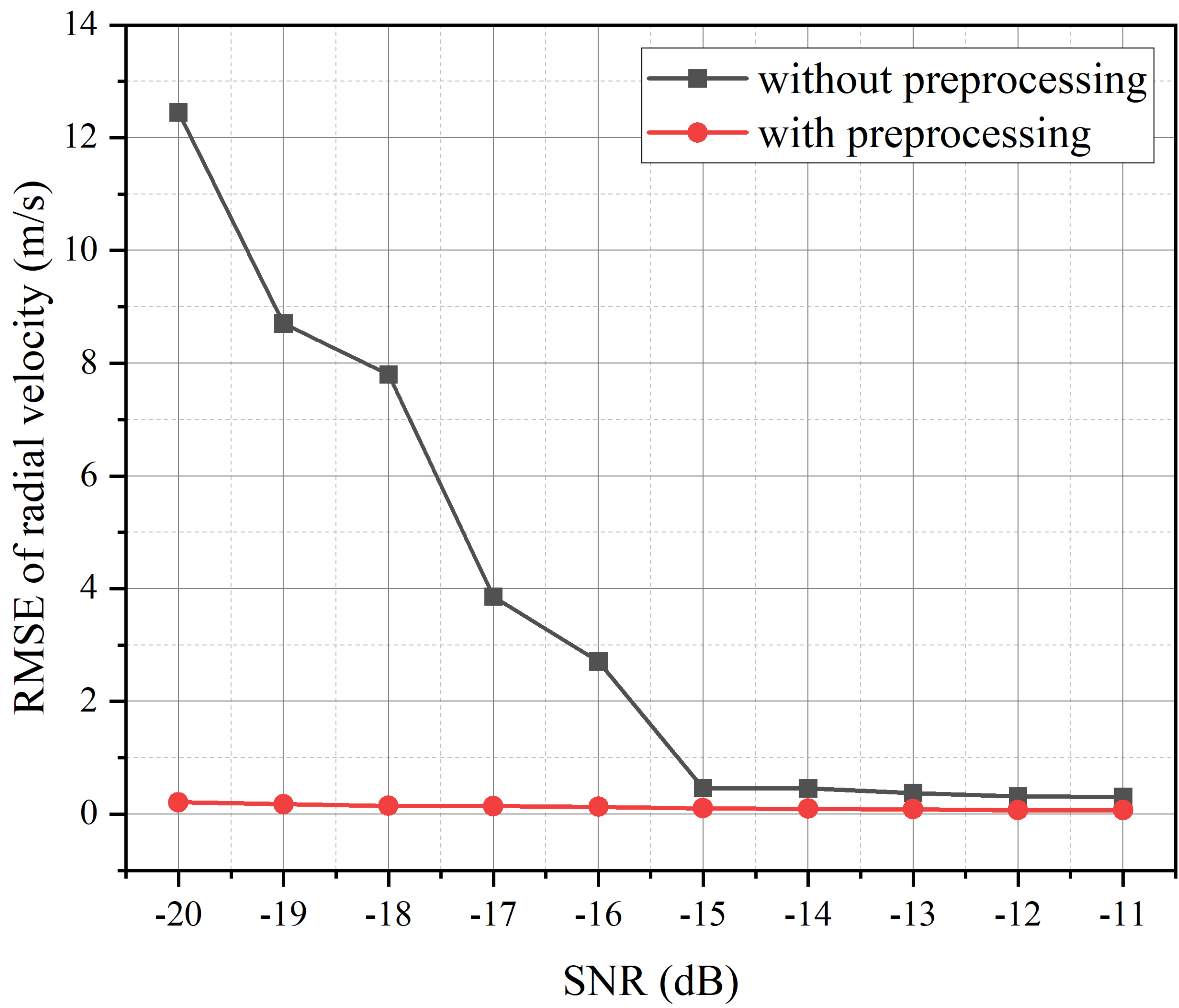}}
	\subfigure[Angle estimation.]{
		\includegraphics[width=0.41\linewidth]{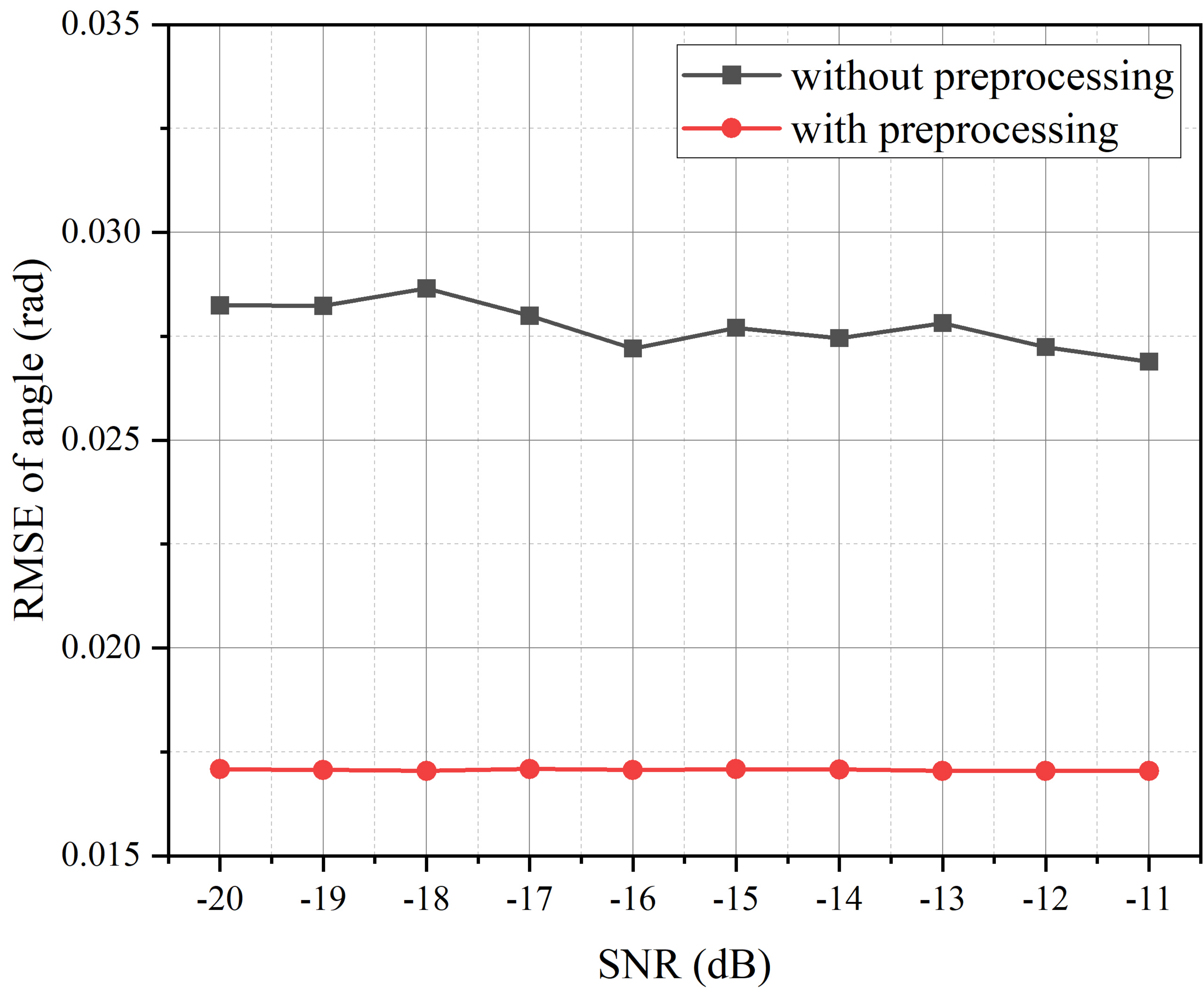}}
	
	\caption{RMSE of single-BS sensing versus SNR with and without preprocessing.}
	\label{fig:Single-BS estimation}
\end{figure}

Fig.~\ref{fig:Single-BS estimation} shows the RMSE results of single-BS sensing with and without preprocessing, with $N_r$ and $N_d$ equal to 100 and 20, respectively. Since the superposition of spectral functions plays a significant role in random noise elimination, the RMSEs of radial distance, radial velocity, and angle estimation are significantly reduced after preprocessing. Furthermore, the performance improvement becomes more evident as the SNR decreases.

\subsection{Simulation Results of Localization}
\begin{figure}[!h]
	\centering
	\includegraphics[width=0.6\linewidth]{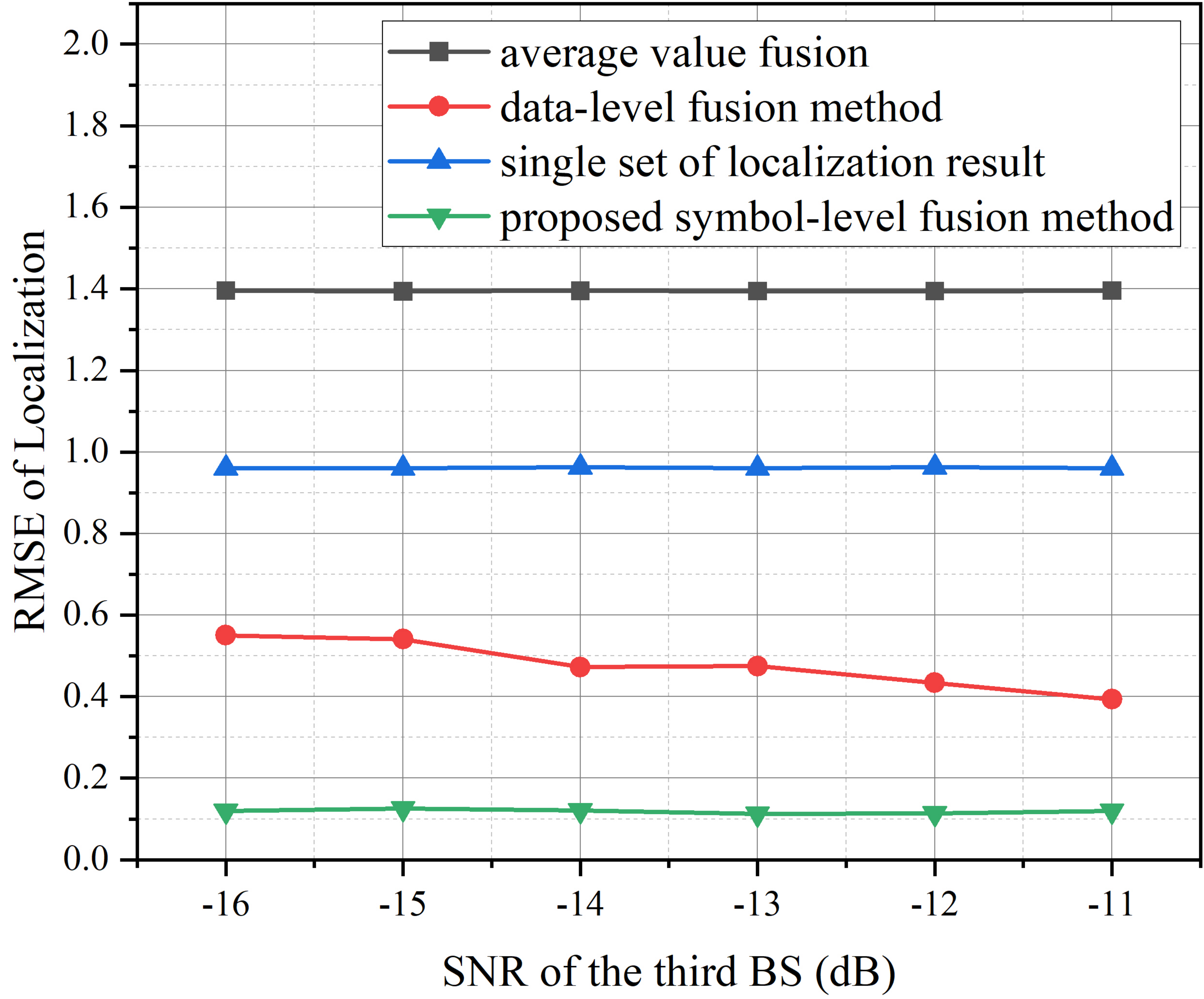}
	\caption{RMSE of localization versus SNR of the third BS.}
	\label{fig:localization RMSE different SNR}
\end{figure}

\begin{figure}[!h]
	\centering
	\includegraphics[width=0.6\linewidth]{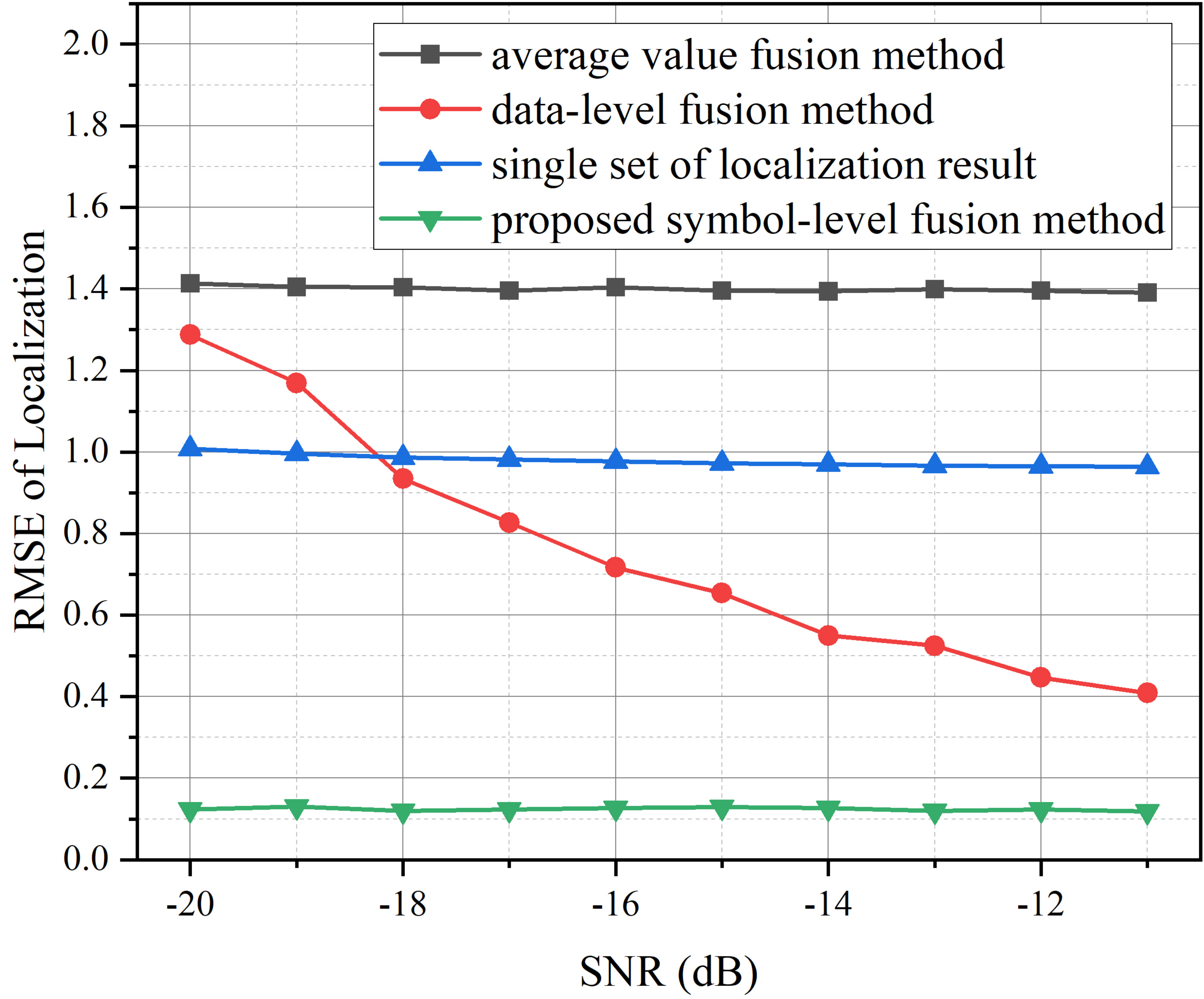}
	\caption{RMSE of localization versus SNR.}
	\label{fig:localization RMSE same SNR}
\end{figure}

The localization results of multiple BSs with different and same SNR scenarios are illustrated in Fig.~\ref{fig:localization RMSE different SNR} and Fig.~\ref{fig:localization RMSE same SNR}. The simulation results are obtained by fusing information from 3 BSs. Due to the characteristics of UAV, the SNR is less than or equal to -10 dB, assuming that the difference in SNR among BSs is not substantial. As for the scenario of BSs with different SNR, the SNR for the first and second BS is set to -10 dB and -12 dB, respectively, and the SNR for the third BS varies from -16 dB to -11 dB. 

Notably, the RMSEs of the proposed symbol-level fusion method are the lowest compared with other localization algorithms. The proposed fusion method also exhibits improved stability across different SNR of BSs due to the stability of a single set of localization result caused by single-BS preprocessing and the ability to reduce information loss compared with the data-level fusion method.

\subsection{Simulation Results of Velocity Estimation}
\begin{figure}[!h]
	\centering
	\includegraphics[width=0.6\linewidth]{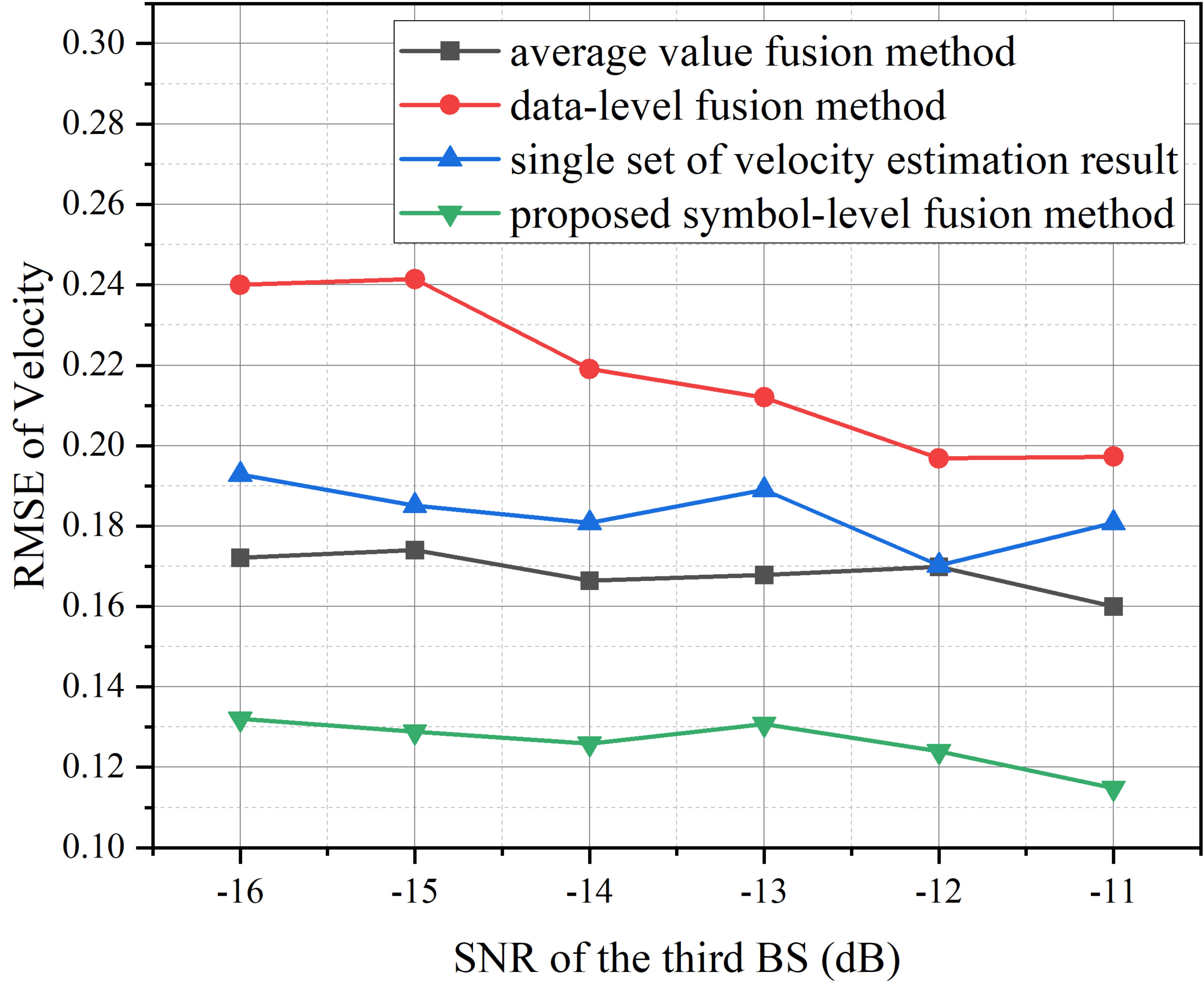}
	\caption{RMSE of velocity estimation versus SNR of the third BS.}
	\label{fig:velocity RMSE different SNR}
\end{figure}

\begin{figure}[!h]
	\centering
	\includegraphics[width=0.6\linewidth]{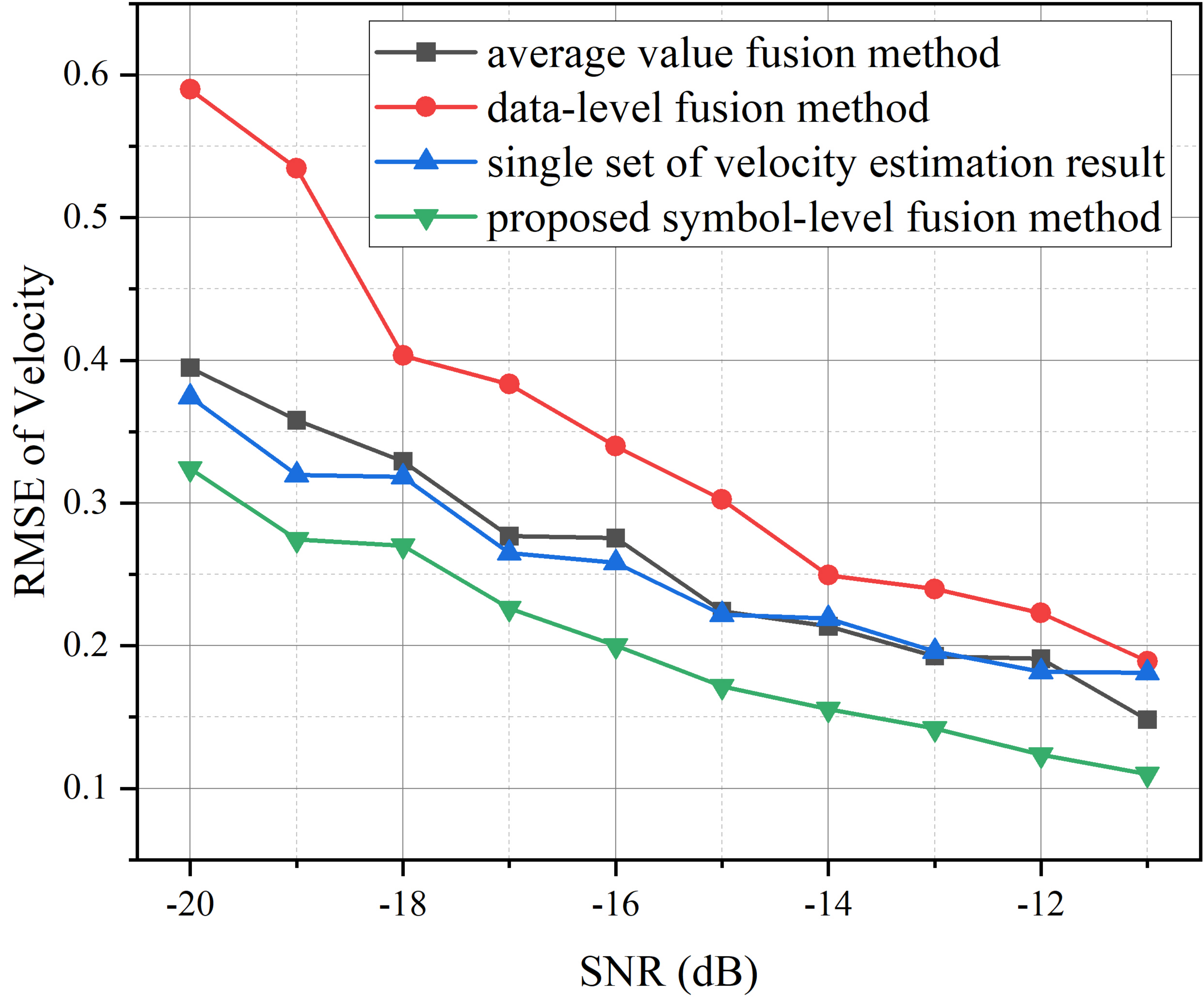}
	\caption{RMSE of velocity estimation versus SNR.}
	\label{fig:velocity RMSE same SNR}
\end{figure}

Fig.~\ref{fig:velocity RMSE different SNR} and Fig.~\ref{fig:velocity RMSE same SNR} demonstrate the velocity estimation results of multiple BSs with different and same SNR scenarios. The simulation results are obtained by fusing information from 4 BSs, with the SNR of the fourth BS set to -10 dB for the scenario of BSs with different SNR.

The proposed symbol-level fusion exhibits superior performance compared with other estimation methods. Nevertheless, it is noteworthy that the velocity error is sensitive to variations in SNR, given that the calculation of the UAV velocity is influenced by information from 3 BSs, the error of which is not as stable as that of single-BS preprocessing. Furthermore, the fluctuation in the RMSEs of different SNR scenario is more pronounced since the SNR difference among BSs results in the fluctuating velocity estimation, further influencing the performance of fusion methods.

\section{Conclusion}
A symbol-level fusion algorithm based on MUSIC algorithm is proposed in this paper, containing a two-step process involving single-BS preprocessing and search of the lattice point with the minimum error. Simulation results demonstrate the evident improvements in accuracy and stability achieved through single-BS preprocessing. Moreover, the symbol-level fusion algorithm has superior performance in UAV localization and velocity estimation compared with the average value fusion method, data-level fusion method and a single set of estimation results. Additionally, symbol-level fusion demonstrates a notable capability for mitigating information loss compared to data-level fusion, further enhancing sensing accuracy. In the future, we will further explore multi-BS cooperative sensing in multi-UAV scenarios to address challenges caused by occlusion issues.

\end{document}